\documentclass[acmsmall]{acmart}

\acmJournal{PACMPL}
\acmVolume{1}
\acmNumber{POPL} 
\acmArticle{1}
\acmYear{2021}
\acmMonth{1}
\acmDOI{} 
\startPage{1}

\setcopyright{none}

\usepackage{stmaryrd}
\usepackage[utf8]{inputenc}
\usepackage{newunicodechar}
\newunicodechar{≥}{\ensuremath{\geq{}}}
\newunicodechar{≤}{\ensuremath{\leq{}}}
\newunicodechar{∨}{\ensuremath{\vee{}}}
\newunicodechar{∧}{\ensuremath{\wedge{}}}
\newunicodechar{♢}{\ensuremath{\diamond{}}}
\newunicodechar{Δ}{\ensuremath{\Delta{}}}
\newunicodechar{∆}{\ensuremath{\Delta{}}}
\newunicodechar{∇}{\ensuremath{\nabla{}}}
\newunicodechar{∀}{\ensuremath{\forall{}}}
\newunicodechar{∈}{\ensuremath{\in{}}}
\newunicodechar{⇆}{\ensuremath{\leftrightarrows{}}}
\newunicodechar{Σ}{\ensuremath{\Sigma{}}}
\newunicodechar{≡}{\ensuremath{\equiv{}}}
\newunicodechar{⊕}{\ensuremath{\oplus{}}}
\newunicodechar{⊛}{\ensuremath{\otimes{}}}
\usepackage{booktabs}   
\usepackage{subcaption} 
\usepackage{listings}
\newcommand{\passthrough}[1]{#1}
\providecommand{\tightlist}{%
  \setlength{\itemsep}{0pt}\setlength{\parskip}{0pt}}

\usepackage{fancyvrb}

\DefineVerbatimEnvironment{Highlighting}{Verbatim}{commandchars=\\\{\}}

\newlength{\cslhangindent}
  {\setlength{\parindent}{0pt}%
  \everypar{\setlength{\hangindent}{\cslhangindent}}\ignorespaces}%
  {\par}

\begin{document}

\title{Curious properties of latency distributions}
  \date{July 10 2020, v2.3}
\title{Curious properties of latency distributions}         

\author{Michał J. Gajda}
\email{mjgajda@migamake.com}

\begin{abstract}
Network latency distributions, their algebra, and use examples.
\end{abstract}



\maketitle

\hypertarget{introduction}{%
\section{Introduction}\label{introduction}}

Here we explain the network simulation context which gave context to
this paper and give key references.

\begin{description}
\tightlist
\item[Capacity insensitive networking]
Most of the network connections on the internet are called
\emph{mice}{[}2{]} for a good reason: they have
\(\text{bandwidth}x\text{latency}\) product of less than 12k, and thus
are largely \emph{insensitive} to the network's capacity. The deciding
factor in performance of these flows is thus \textbf{latency}, hence we
ignore capacity limitation for the remainder of this paper.
\item[Packet loss modelling using improper CDFs]
In order to accurately simulate capacity-insensitive network
miniprotocols we formally define network latency distribution as
improper CDF (cumulative distribution function) of arrived messages over
time. We call it improper CDF, because it does not end at 100\%: some
messages can be lost.
\item[Time-limited model]
For practical purposes we ignore answers that are delivered after
certain deadline: that is network connection timeout.
\end{description}

Starting with description of its apparent properties, we identify their
mathematical definitions, and ultimately arrive at algebra of ΔQ {[}3{]}
with basic operations that correspond to abstract interpretations of
network miniprotocols{[}8{]}.

This allows us to use objects from single algebraic body to describe
behaviour of entire protocols as improper CDFs.

\hypertarget{related-work}{%
\subsection{Related work}\label{related-work}}

Then we discuss expansion of the concept to get most sensitive metrics
of protocol and network robustness{[}4{]}. However instead of heuristic
measure like effective graph resistance {[}5{]}, we use logically
justified measure derived from the actual behaviour of the network.

This is similar to \emph{network calculus}{[}6{]} but uses simpler
methods and uses more logical description with improper latency
distribution functions, instead of \emph{max-plus} and \emph{min-plus}
algebras.\footnote{We describe how it generalizes these \emph{max-plus}
  and \emph{min-plus} algebras later.} Basic operations ∧ and ∨ are
similar to last-to-finish and first-to-finish synchronizations {[}3{]}.

This approach allows us to use \emph{abstract interpretation}{[}8{]} of
computer program to get its latency distribution, or a single execution
to approximate latency distribution assuming the same loss profile of
packets.

\hypertarget{preliminaries}{%
\section{Preliminaries}\label{preliminaries}}

\hypertarget{nulls-and-units-of-multiplication}{%
\subsubsection{Nulls and units of
multiplication}\label{nulls-and-units-of-multiplication}}

We will be interested in null and unit of a single multiplication for
each modulus we will consider:

\begin{lstlisting}[language=Haskell]
class Unit a where
  unitE :: a
class Null a where
  nullE :: a
instance Unit Int where
  unitE = 1
instance Null Int where
  nullE = 0
\end{lstlisting}

\hypertarget{discrete-delay}{%
\subsection{Discrete delay}\label{discrete-delay}}

Discrete delays are defined as:

\begin{lstlisting}[language=Haskell]
newtype Delay = Delay { unDelay :: Int }
\end{lstlisting}

For ease of implementation, we express each function as a series of
values for \protect\hyperlink{delay}{discrete delays}. First value is
for \emph{no delay}. We define \(start \in{}\mathcal{T}\) as smallest
\passthrough{\lstinline!Delay!} (no delay).

\begin{lstlisting}[language=Haskell]
start :: Delay
start = Delay 0
\end{lstlisting}

\hypertarget{series}{%
\subsection{Power series representing distributions}\label{series}}

We follow {[}7{]} exposition of power series, but use use finite series
and shortcut evaluation:

\begin{lstlisting}[language=Haskell]
newtype Series a = Series { unSeries :: [a] }
  deriving (Eq, Show, Read, Functor, Foldable,
            Applicative, Semigroup)
\end{lstlisting}

To have a precise discrete approximation of distributions, we encode
them as a series.

Finite series can be representened by a generating function \(F_t\):

\[ F(t)=f_0*t^0+f_1*t^1+f_2*t^2+...+f_n*t^n \]

Which is represented by the Haskell data structure:

\begin{lstlisting}[language=Haskell]
f_t = Series [a0, a1, .., an]
\end{lstlisting}

\hypertarget{differential-encoding-and-cumulative-sum}{%
\subsubsection{Differential encoding and cumulative
sum}\label{differential-encoding-and-cumulative-sum}}

For any probability distribution, we need a notion of integration, that
coverts probability distribution function (PDF) into cumulative
distribution function (CDF).

\textbf{Cumulative sums} computes sums of 1..n-th term of the series:

\begin{lstlisting}[language=Haskell]
cumsum :: Num a => Series a -> Series a
cumsum = Series . tail . scanl (+) 0 . unSeries
\end{lstlisting}

After defining our integration operator, it is not time for its inverse.
\textbf{Differential encoding} is lossless correspondent of discrete
differences,\footnote{Usually called \emph{finite difference operators}.}
but with first coefficient being copied. (Just like there was a zero
before each series, so that we always preserve information about the
first term.) This is \emph{backward antidifference}\footnote{Antidifference
  is an inverse of finite difference operator. Backwards difference
  subtracts immediate \emph{predecessor} from a successor term in the
  series.} as defined by {[}11{]}.

That makes it an inverse of \passthrough{\lstinline!cumsum!}. It is
\emph{backward finite difference operator}, as defined by {[}10{]} .

\begin{lstlisting}[language=Haskell]
diffEnc :: Num a => Series a -> Series a
diffEnc (Series []) = Series []
diffEnc (Series s ) = Series $
  head s : zipWith (-) (tail s) s
\end{lstlisting}

So that \passthrough{\lstinline!diffEnc!} of CDF will get PDF, and
\passthrough{\lstinline!cumsum!} of PDF will get CDF.

Since we are only interested in answers delivered before certain
deadline, we will sometimes cut series at a given index:

\begin{lstlisting}[language=Haskell]
cut :: Delay -> Series a -> Series a
cut (Delay t) (Series s) = Series (take t s)

instance IsList (Series a) where
  type Item (Series a) = a
  fromList          = Series
  toList (Series s) = s
\end{lstlisting}

Series enjoy few different multiplication operators.

Simplest is scalar multiplication:

\begin{lstlisting}[language=Haskell]
infixl 7 .* -- same precedence as *
(.*):: Num a => a -> Series a -> Series a
c .* (Series (f:fs)) =
      Series (c*f
             :unSeries ( c.* Series fs))
_ .* (Series []    ) = Series []
\end{lstlisting}

\[ F(t)=f_0*t^0+f_1*t^1+f_2*t^2+...+f_n*t^n \]

Second multiplication is convolution, which most commonly used on
distributions: \[F(t)⊛G(t)=\Sigma_{\tau=0}^t x^t*f(\tau)*g(t-\tau) \]
Wikipedia's definition:
\[(f ⊛ g)(t) \triangleq\ \int_{-\infty}^\infty f(\tau) g(t - \tau) \, d\tau.\]
Distribution is from \(0\) to \(+\infty\): Wikipedia's definition:

\begin{enumerate}
\def\labelenumi{\arabic{enumi}.}
\item
  First we fix the boundaries of integration:
  \[(f ⊛ g)(t) \triangleq\ \int_{0}^\infty f(\tau) g(t - \tau) \, d\tau.\]
  (Assuming \(f(t)=g(t)=0\) when \(t<0\).)
\item
  Now we change to discrete form:
\end{enumerate}

\[(f ⊛ g)(t) \triangleq\ \Sigma_{0}^\infty f(\tau) g(t - \tau)\]

\begin{enumerate}
\def\labelenumi{\arabic{enumi}.}
\setcounter{enumi}{2}
\tightlist
\item
  Now we notice that we approximate functions up to term \(n\):
  \[(f ⊛ g)(t) \triangleq\ \Sigma_{0}^{n} f_{\tau} g_{t - \tau}.\]
\end{enumerate}

Resulting in convolution: \[F(t)⊛G(t)=Σ_{\tau{}=0}^t x^t*f(\tau)*g(t-
\tau) \]

\begin{lstlisting}[language=Haskell]
infixl 7 `convolve` -- like multiplication
convolve :: Num    a
         => Series a
         -> Series a
         -> Series a
Series            (f:fs)
  `convolve`
     gg@(Series (g:gs)) =
  Series
    (f*g :
      unSeries (f .* Series gs +
               (Series fs `convolve` gg)))
Series []     `convolve` _         = Series []
_             `convolve` Series [] = Series []
\end{lstlisting}

Elementwise multiplication, assuming missing terms are zero.

\begin{lstlisting}[language=Haskell]
(.*.) :: Num    a
      => Series a
      -> Series a
      -> Series a
Series a .*. Series b = Series (zipWith (*) a b)
\end{lstlisting}

Since we use finite series, we need to extend their length when
operation is done on series of different length.

Note that for emphasis, we also allow convolution with arbitrary
addition and multiplication:

\begin{lstlisting}[language=Haskell]
convolve_ :: (      a ->        a ->        a)
          -> (      a ->        a ->        a)
          -> Series a -> Series a -> Series a
convolve_ (+) (*) (Series (f:fs)) gg@(Series (g:gs)) =
  Series
    (f * g :
      zipWithExpanding
        (+)
        (f .* gs)
        (unSeries (convolve_ (+) (*)
                             (Series fs) gg)))
  where
    a .* bs = (a*) <$> bs
convolve_ _ _ (Series [])  _          = Series []
convolve_ _ _  _          (Series []) = Series []
\end{lstlisting}

\hypertarget{expanding-two-series-to-the-same-length}{%
\subsection{Expanding two series to the same
length}\label{expanding-two-series-to-the-same-length}}

We need a variant of \passthrough{\lstinline!zipWith!} that assumes that
shorter list is expanded with unit of the operation given as argument:

\begin{lstlisting}[language=Haskell]
zipWithExpanding :: (a  ->  a  ->  a)
                 -> [a] -> [a] -> [a]
zipWithExpanding f = go
  where
    go    []     ys  = ys -- unit `f` y    == y
    go    xs     []  = xs -- x    `f` unit == x
    go (x:xs) (y:ys) = (x `f` y):go xs ys
\end{lstlisting}

Here we use extension by a given element \passthrough{\lstinline!e!},
which is 0 for normal series, or 1 for complement series.

Extend both series to the same length with placeholder zeros. Needed for
safe use of complement-based operations.

\begin{lstlisting}[language=Haskell]
extendToSameLength e (Series a, Series b) =
    (Series resultA, Series resultB)
  where
    (resultA, resultB) = go a b
    go  []       []  = (    [] ,    [] )
    go (b:bs) (c:cs) = (  b:bs',  c:cs')
      where
        ~(bs', cs') = go bs cs
    go (b:bs)    []  = (  b:bs, e  :cs')
      where
        ~(bs', cs') = go bs []
    go    []  (c:cs) = (e  :bs',  c:cs )
      where
        ~(bs', _  ) = go [] cs
\end{lstlisting}

In a rare case (CDFs) we might also prolong by the length of the last
entry:

We will sometimes want to extend both series to the same length with
placeholder of last element.

\begin{lstlisting}[language=Haskell]
extendToSameLength' (Series a, Series b) =
    (Series resultA, Series resultB)
  where
    (resultA, resultB) = go a b
    go  []       []  = (    [] ,    [] )
    go  [b]     [c]  = (   [b] ,   [c] )
    go (b:bs)   [c]  = (  b:bs,   c:cs')
      where
        ~(bs', cs') = go bs [c]
    go   [b]  (c:cs) = (b  :bs',  c:cs )
      where
        ~(bs', _  ) = go [b] cs
    go (b:bs) (c:cs) = (  b:bs',  c:cs')
      where
        ~(bs', cs') = go bs cs
\end{lstlisting}

\hypertarget{series-modulus}{%
\subsection{Series modulus}\label{series-modulus}}

We can present an instance of number class for
\passthrough{\lstinline!Series!}:

\begin{lstlisting}[language=Haskell]
instance Num a => Num (Series a) where
  Series a + Series b = Series
    (zipWithExpanding (+) a b)
  (*)         = convolve
  abs         = fmap abs
  signum      = fmap signum
  fromInteger = seriesFromInteger
  negate      = fmap negate
\end{lstlisting}

Note that we do not know yet how to define
\passthrough{\lstinline!fromInteger!} function. Certainly we would like
to define null and unit (neutral element) of convolution, but it is not
clear what to do about the others:

\begin{lstlisting}[language=Haskell]
seriesFromInteger 0     = nullE
seriesFromInteger 1     = unitE
seriesFromInteger other =
  error $ "Do not use fromInteger "
       <> show other <> " to get Series!"
\end{lstlisting}

Given a \emph{unit} and \emph{null} elements, we can give unit and null
element of a \passthrough{\lstinline!Series!}.\footnote{Note that we in
  this context we are mainly interested in null and unit of
  multiplication.}

\begin{lstlisting}[language=Haskell]
instance Unit a => Unit (Series a) where
  unitE = [unitE]

instance Null a => Null (Series a) where
  nullE = [nullE]
\end{lstlisting}

We may be using \passthrough{\lstinline!Series!} of floating point
values that are inherently approximate.

In this case, we should not ever use equality, but rather a similarity
metric that we can generate from the similar metric on values:

\begin{lstlisting}[language=Haskell]
instance Real           a
      => Metric (Series a) where
  a `distance` b = sqrt
                 $ realToFrac
                 $ sum
                 $ fmap square
                 $ a-b
  similarityThreshold = 0.001

square x = x*x
\end{lstlisting}

Note that generous similarity threshold of
\passthrough{\lstinline!0.001!} is due to limited number of simulations
we can quickly run when checking distributions in the unit tests (10k by
default).

For a \passthrough{\lstinline!Series!} of of objects having complement,
there is also well established definition:

\begin{lstlisting}[language=Haskell]
instance Complement         a
      => Complement (Series a) where
  complement = fmap complement
\end{lstlisting}

\hypertarget{square-matrices-of-declared-size}{%
\subsection*{Square matrices of declared
size}\label{square-matrices-of-declared-size}}
\addcontentsline{toc}{subsection}{Square matrices of declared size}

This is a simple description of square matrices with fixed
size.\footnote{Note that we considered using
  \passthrough{\lstinline!matrix-static!}, but it does not have typesafe
  indexing.} First we need natural indices that are no larger than
\(n\):

\begin{lstlisting}[language=Haskell]
newtype UpTo (n::Nat) = UpTo { _unUpTo :: Natural }
  deriving (Eq, Ord, Num, Typeable)
\end{lstlisting}

The only purpose of safe indices is to enumerate them:

\begin{lstlisting}[language=Haskell]
allUpTo :: KnownNat n => [UpTo n]
\end{lstlisting}

Armed with safe indices, we can define square matrices:

\begin{lstlisting}[language=Haskell]
newtype SMatrix (n::Nat) a =
    SMatrix { unSMatrix :: DM.Matrix a }
  deriving (Show, Eq, Functor, Applicative
           ,Foldable, Traversable,  Typeable, Generic)
\end{lstlisting}

\begin{lstlisting}[language=Haskell]
size :: KnownNat n => SMatrix n a -> Int
size (s :: SMatrix n a)= intVal (Proxy :: Proxy n)
\end{lstlisting}

We also need to identity and null matrices (for multiplication):

\begin{lstlisting}[language=Haskell]
instance (KnownNat      n
         ,Null            a)
      =>  Null (SMatrix n a) where
  nullE = sMatrix Proxy (\_ -> nullE)

instance (KnownNat      n
         ,Null            a
         ,Unit            a)
      =>  Unit (SMatrix n a) where
  unitE = sMatrix Proxy elt
    where
      elt (i,j) | i == j = unitE
      elt (i,j)          = nullE
\end{lstlisting}

Definition of parametrized matrix multiplication is standard, so we can
test it over other objects with defined multiplication and addition-like
operators.

\begin{lstlisting}[language=Haskell]
sMatMult ::  KnownNat n
         => (a -> a -> a) -- ^ addition
         -> (a -> a -> a) -- ^ multiplication
         ->  SMatrix  n a
         ->  SMatrix  n a
         ->  SMatrix  n a
sMatMult add mul a1 (a2 :: SMatrix n a) =
    sMatrix (Proxy :: Proxy n) gen
  where
    gen ::  KnownNat n
        => (UpTo n, UpTo n)
        ->  a
    gen (i,j) = foldr1 add
              [ (a1 ! (i,k)) `mul` (a2 ! (k,j))
              | k <- allUpTo' i ]
\end{lstlisting}

Note that to measure convergence of the process, we need a notion of
distance between two matrices.

Matrix addition for testing:

\begin{lstlisting}[language=Haskell]
(|+|) :: (Num        a
         ,KnownNat n  )
      => SMatrix   n a
      -> SMatrix   n a
      -> SMatrix   n a
a |+| b = (+) <$> a <*> b
\end{lstlisting}

One might also want to iterate over rows or columns in the matrix:

\begin{lstlisting}[language=Haskell]
rows, columns :: KnownNat n
              => SMatrix n a -> [[a]]
rows    sm = [[sm ! (i,j) | j<-allUpTo ] | i<-allUpTo]
columns sm = [[sm ! (i,j) | i<-allUpTo ] | j<-allUpTo]
\end{lstlisting}

\hypertarget{latency-distributions}{%
\section{Latency distributions}\label{latency-distributions}}

\hypertarget{introducing-improper-cdf}{%
\subsection{Introducing improper CDF}\label{introducing-improper-cdf}}

To define our key object, lets imagine a single network connection. In
this document, we ignore capacity-related issues. So \(∆Q(t)\) is
\emph{improper cumulative distribution function} of event arriving at
some point of time:

\begin{figure}
\centering
\includegraphics[width=0.5\textwidth,height=0.25\textheight]{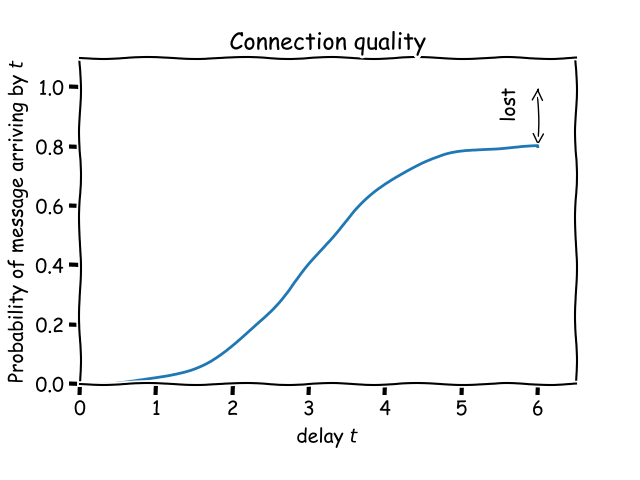}
\caption{Completion rate against deadline}
\end{figure}

\begin{figure}
\centering
\includegraphics[width=0.5\textwidth,height=0.25\textheight]{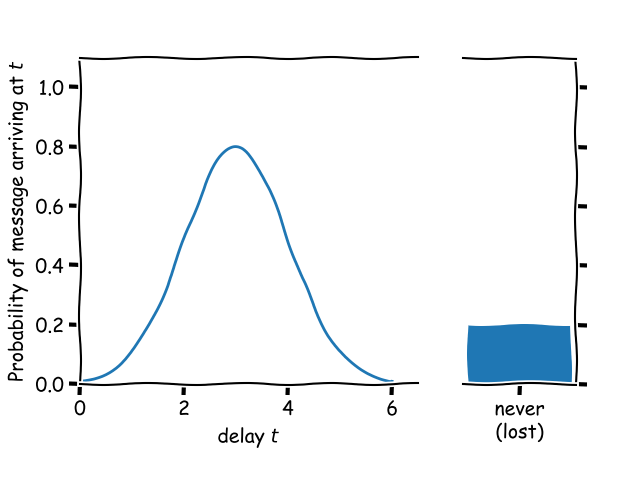}
\caption{Latency distribution}
\end{figure}

For the sake of practicality, we also mark a \emph{deadline} as the last
possible moment when we still care about messages. (Afterwards, we drop
them, just like TCP timeout works.)

For example, when only \(0.99\) of messages arrive at all within desired
time \(t\), and we silently drop those that arrive later.

For each distribution, we will define \emph{deadline} formally as
\(d(t)=\text{maxarg}_{t}(ΔQ(t))\) or such \(t\) for which our improper
CDF reaches maximum value. We also define \emph{ultimate arrival
probability} formally as \(a_{u}(ΔQ)=\max(ΔQ)\). Our improper CDFs are
assumed to be always defined within finite subrange of delays, starting
at \(0\). Ultimate arrival probability allows us to compare attenuation
between two links.

In the following we define domain of \emph{arrival probabilities} as
\(\mathcal{A}\in[0,1.0]\), which is
\protect\hyperlink{probability}{probability}.

We also define domain of \emph{time} or \emph{delays} as
\(\mathcal{T}\). We also call a domain of \(ΔQ\) functions as
\(\mathcal{Q}=(\mathcal{T}\rightarrow{}\mathcal{A})\).

Below is Haskell specification of this datatype:

\begin{lstlisting}[language=Haskell]
newtype LatencyDistribution a =
  LatencyDistribution { pdf :: Series a }
\end{lstlisting}

The representation above holds PDF (probability density function). Its
cumulative distribution function can be trivially computed with running
sum:

\begin{lstlisting}[language=Haskell]
cdf :: Num                 a
    => LatencyDistribution a
    -> Series              a
cdf = cumsum . pdf
\end{lstlisting}

Since it is common use complement of CDF, we can have accessor for this
one too:

\begin{lstlisting}[language=Haskell]
complementCDF :: Probability         a
              => LatencyDistribution a
              -> Series              a
complementCDF  = complement . cumsum . pdf
\end{lstlisting}

\hypertarget{canonical-form}{%
\subsection{Canonical form}\label{canonical-form}}

Sometimes we need to convert possibly improper
\passthrough{\lstinline!LatencyDistribution!} into its canonical
representation.

Here we define \emph{canonical}
\passthrough{\lstinline!LatencyDistribution!} when (i) it is a valid
improper probability distribution so sum does not go over \(1.0\); (ii)
it does not contain trailing zeros after the first element (which are
redundant). This definition assumes we have a finite series, and assures
that any distribution has a unique representation.

\begin{lstlisting}[language=Haskell]
canonicalizeLD :: Probability         a
               => LatencyDistribution a
               -> LatencyDistribution a
canonicalizeLD = LatencyDistribution
               . Series
               . assureAtLeastOneElement
               . dropTrailingZeros
               . cutWhenSumOverOne 0.0
               . unSeries
               . pdf
  where
    cutWhenSumOverOne aSum []           = []
    cutWhenSumOverOne aSum (x:xs)
                           | aSum+x>1.0 = [1.0-aSum]
    cutWhenSumOverOne aSum (x:xs)       =
        x:cutWhenSumOverOne (aSum+x) xs
    assureAtLeastOneElement []          = [0.0]
    assureAtLeastOneElement other       = other
    dropTrailingZeros                   = reverse
                    . dropWhile (==0.0) . reverse
\end{lstlisting}

\hypertarget{construction-from-pdf-and-cdf}{%
\subsection{Construction from PDF and
CDF}\label{construction-from-pdf-and-cdf}}

We use \passthrough{\lstinline!canonicalizeLD!} to make sure that every
distribution is kept in canonical form (explained below), we might also
want to make constructors that create
\passthrough{\lstinline!LatencyDistribution!} from a series that
represents PDF or CDF:

\begin{lstlisting}[language=Haskell]
fromPDF :: Probability         a
        => Series              a
        -> LatencyDistribution a
fromPDF = canonicalizeLD . LatencyDistribution
\end{lstlisting}

To create LatencyDistribution from CDF we need
\passthrough{\lstinline!diffEnc!} (\emph{differential encoding} or
\emph{backward finite difference operator} from
\protect\hyperlink{series}{\passthrough{\lstinline!Series!} module}):

\begin{lstlisting}[language=Haskell]
fromCDF :: Probability         a
        => Series              a
        -> LatencyDistribution a
fromCDF = fromPDF . diffEnc
\end{lstlisting}

Similar we can create \passthrough{\lstinline!LatencyDistribution!} from
complement of CDF:

\begin{lstlisting}[language=Haskell]
fromComplementOfCDF :: Probability         a
                    => Series              a
                    -> LatencyDistribution a
fromComplementOfCDF  = fromCDF . complement
\end{lstlisting}

\hypertarget{intuitive-properties-of-latency-distributions}{%
\subsection{Intuitive properties of latency
distributions}\label{intuitive-properties-of-latency-distributions}}

\begin{enumerate}
\def\labelenumi{\arabic{enumi}.}
\item
  We can define few \emph{linear} operators on ΔQ (for exact definition,
  see next section):

  A. Stretching in time -- ignored in here.

  B. Delaying by \(t\) -- composition with \(\text{wait}\):
  \[wait(t)=f(t)= \begin{cases} 0 & \text{for } t<t_d \\ 1.0 & \text{for } t=t_d \end{cases}\]

  C. Scaling arrival probability -- in other words,
  \(\text{attenuation}\).
\item
  We distinguish special distribution that represents \emph{null delay}
  or \emph{neutral element} of sequential composition (\(\fatsemi{}\) or
  \passthrough{\lstinline!afterLD!}), where we pass every message with
  no delay: \[preserved(1)=wait(0)=1_{\mathcal{Q}}\]
\item
  We can say that one \(ΔQ\) no worse than the other, when it is
  improper CDF values never less than the other after making it fit a
  common domain: \[
  ΔQ_1≥_QΔQ_2 ≡ ∀t. X(ΔQ_1)(t)≥X(ΔQ_2)(t)
  \] Here assuming \(X(ΔQ)\) defined: \[
  X(ΔQ)(t)≡\begin{cases}ΔQ(t)     & \text{for } t≤d(ΔQ)\\
                     ΔQ(d(ΔQ)) & \text{otherwise}
        \end{cases}
  \]
\end{enumerate}

\hypertarget{basic-operations-on-ux3b4q}{%
\subsection{Basic operations on ΔQ}\label{basic-operations-on-ux3b4q}}

To model connections of multiple nodes, and network protocols we need
two basic operations: sequential and parallel composition.

Interestingly both of these operations form an associative-commutative
monoids (with unit that changes nothing) with null element (zero that
nullifies the other), however their null and neutral elements swap
places.

\begin{enumerate}
\def\labelenumi{\arabic{enumi}.}
\item
  Sequential composition~\(\fatsemi{}\) or
  \passthrough{\lstinline!afterLD!}:\footnote{Sometimes named
    \(\mathbf{;}\).} given \(ΔQ_1(t)\) and \(ΔQ_2(t)\) of two different
  connections, we should be able to compute the latency function for
  routing the message through pair of these connections in sequence:
  \[ΔQ_1(t)\fatsemi{}ΔQ_2(t)\].

  \begin{itemize}
  \tightlist
  \item
    \emph{associative}:
    \[ΔQ_1(t) \fatsemi{} [ΔQ_2(t)\fatsemi{}ΔQ_3(t)]=[ΔQ_1(t)\fatsemi{}ΔQ_2(t)]\fatsemi{}ΔQ_3(t)\]
  \item
    \emph{commutative}
    \[ΔQ_1(t) \fatsemi{} ΔQ_2(t)=ΔQ_2(t) \fatsemi{} ΔQ_1(t)\]
  \item
    \emph{neutral element} is \(1_{\mathcal{Q}}\) or
    \passthrough{\lstinline!noDelay!}, so:
    \[ΔQ(t) \fatsemi{} 1_{\mathcal{Q}}=1_{\mathcal{Q}} \fatsemi{} ΔQ(t)=ΔQ(t)\]
  \item
    \emph{null element} is \(0_{\mathcal{Q}}\) or
    \passthrough{\lstinline!allLost!}, so:
    \[ΔQ(t) \fatsemi{} 0_{\mathcal{Q}}=0_{\mathcal{Q}} \fatsemi{} ΔQ(t)=0_{\mathcal{Q}}\]
  \end{itemize}
\end{enumerate}

\begin{lstlisting}[language=Haskell]
afterLD :: Probability a
        => LatencyDistribution a
        -> LatencyDistribution a
        -> LatencyDistribution a
rd1 `afterLD` rd2 = fromPDF
                  $ pdf rd1 `convolve` pdf rd2
\end{lstlisting}

\begin{enumerate}
\def\labelenumi{\arabic{enumi}.}
\setcounter{enumi}{1}
\tightlist
\item
  Alternative selection ∨: given \(ΔQ_1(t)\) and \(ΔQ_2(t)\) of two
  different connections, we should be able to compute the latency
  function for routing the message through pair of these connections in
  parallel:
\end{enumerate}

\[ΔQ_1(t)\mathbf{∨} ΔQ_2(t)\] * \emph{associative}:
\[ΔQ_1(t)∨[ΔQ_2(t)∨ΔQ_3(t)]=[ΔQ_1(t)∨ΔQ_2(t)]∨ΔQ_3(t)\] *
\emph{commutative} \[ΔQ_1(t)∨ΔQ_2(t)=ΔQ_2(t)∨ΔQ_1(t)\] * \emph{neutral
element} is \(0_{\mathcal{Q}}\) or \passthrough{\lstinline!allLost!},
so: \[ΔQ(t) ∨ 0_{\mathcal{Q}}=0_{\mathcal{Q}}∨ΔQ(t)=ΔQ(t)\] * \emph{null
element} of \passthrough{\lstinline!firstToFinish!} is \(1_Q\):
\[ΔQ(t) ∨ 1_{\mathcal{Q}}=1_{\mathcal{Q}}∨ΔQ(t)=1_{\mathcal{Q}}\] *
\emph{monotonically increasing}: \[ΔQ_1(t) ∨ ΔQ_2(t)≥ΔQ_1(t)\]

Here is the Haskell code for naive definition of these two operations:
We can also introduce alternative of two completion distributions. It
corresponds to a an event that is earliest possible conclusion of one of
two alternative events.

That can be easily expressed with improper cumulative distribution
functions:
\[ \text{P}_{min(a,b)}(x \leq t)= 1-(1-P_a(x\leq t)) * (1-P_b(x\leq t))\]
That is, event \(min(a,b)\) occured when \(t<a\) or \(t<b\), when:

\begin{itemize}
\tightlist
\item
  it \textbf{did not occur} (top complement: \(1-...\)), that:

  \begin{itemize}
  \tightlist
  \item
    either \(a\) did \textbf{not} occur \(1-P_a(x≤t)\),
  \item
    and \(b\) did \textbf{not} occur \(1-P_b(x≤t)\):
  \end{itemize}

\begin{lstlisting}[language=Haskell]
firstToFinishLD :: Probability a
              => LatencyDistribution a
              -> LatencyDistribution a
              -> LatencyDistribution a
rd1 `firstToFinishLD` rd2 =
  fromComplementOfCDF   $
    complementCDF rd1' .*.
    complementCDF rd2'
where
  (rd1',
   rd2') = extendToSameLengthLD (rd1
                                ,rd2)
\end{lstlisting}
\end{itemize}

Notes:

\begin{enumerate}
\def\labelenumi{\arabic{enumi}.}
\tightlist
\item
  Since we model this with finite discrete series, we first need to
  extend them to the same length.
\item
  Using the fact that \passthrough{\lstinline!cumsum!} is discrete
  correspondent of integration, and \passthrough{\lstinline!diffEnc!} is
  its direct inverse (\emph{backward finite difference}), we can try to
  differentiate this to get PDF directly:
  \[ \begin{array}{lcr} P_a(x≤t) & = & Σ_{x=0}^{t}P_a(x) \\
  ∇\left(Σ_0^{t}P_{a}(x)dx\right) & = & P_a(t) \\
  \end{array}\] In continuous domain, one can also differentiate both
  sides of the equation above to check if we can save computations by
  computing PDF directly.
\end{enumerate}

Unfortunately, that means that instead of 2x cumulative sum operations,
1x elementwise multiplication, and 1x differential encoding operation,
we still need to perform the same 2x cumulative sums, and 3x pointwise
additions and 3x pointwise multiplications, and two complements.

So we get an equation that is less obviously correct, and more
computationally expensive.

Code would look like:

\begin{lstlisting}[language=Haskell]
rd1 `firstToFinishLD` rd2 = canonicalizeLD $
  LatencyDistribution {
     pdf      = rd1' .*. complementCDF rd2'
              + rd2' .*. complementCDF rd1'
              + rd1' .*. rd2'              }
  where
    (rd1', rd2') = extendToSameLengthLD (rd1, rd2)
    complement :: Series Probability
               -> Series Probability
    complement = fmap (1.0-)
\end{lstlisting}

\emph{Note that \passthrough{\lstinline!complement!} above will be
correct only if both lists are of the same length.}

In order to use this approach in here, we need to prove that
\passthrough{\lstinline!cumsum!} and \passthrough{\lstinline!diffEnc!}
correspond to integration, and differentiation operators for discrete
time domain.

Now let's define neutral elements of both operations above:

\begin{lstlisting}[language=Haskell]
preserved :: Probability         a
          =>                     a
          -> LatencyDistribution a
preserved a = LatencyDistribution {
                pdf = Series [a]  }

allLostLD, noDelayLD :: Probability         a
                     => LatencyDistribution a
allLostLD = preserved 0.0
noDelayLD = preserved 1.0
\end{lstlisting}

Here:

\begin{itemize}
\tightlist
\item
  \passthrough{\lstinline!allLost!} indicates that no message arrives
  ever through this connection
\item
  \passthrough{\lstinline!noDelay!} indicates that the all messages
  always arrive without delay
\end{itemize}

\begin{enumerate}
\def\labelenumi{\arabic{enumi}.}
\setcounter{enumi}{2}
\tightlist
\item
  Conjunction of two different actions simultaneously completed in
  parallel, and waits until they both are:
  \[ \text{P}_{max(a,b)}(x \leq t)= P_a(x\leq t) * P_b(x\leq t)\]
\end{enumerate}

\begin{lstlisting}[language=Haskell]
lastToFinishLD :: Probability a
               => LatencyDistribution a
               -> LatencyDistribution a
               -> LatencyDistribution a
rd1 `lastToFinishLD` rd2 = fromCDF
                         $ cdf rd1' .*. cdf rd2'
  where
    (rd1', rd2') = extendToSameLengthLD (rd1, rd2)
\end{lstlisting}

(Attempt to differentiate these by parts also leads to more complex
equation:
\passthrough{\lstinline!rd1 .*. cumsum rd2 + rd2 .*. cumsum rd1!}.)

Now we can make an abstract interpretation of protocol code to derive
corresponding improper CDF of message arrival.

It is also:

\begin{itemize}
\tightlist
\item
  commutative
\item
  associative
\item
  with neutral element of noDelay
\end{itemize}

\begin{enumerate}
\def\labelenumi{\arabic{enumi}.}
\setcounter{enumi}{3}
\tightlist
\item
  Failover \(A<t>B\) when action is attempted for a fixed period of time
  \(t\), and if it does not complete in this time, the other action is
  attempted:
\end{enumerate}

\begin{lstlisting}[language=Haskell]
failover deadline rdTry rdCatch = fromPDF $
    initial <> fmap (remainder*) (pdf rdCatch)
  where
    initial = cut deadline $ pdf rdTry
    remainder = 1 - sum initial
\end{lstlisting}

Algebraic properties of this operator are clear:

\begin{itemize}
\item
  Certain failure converts deadline into delay:
  \[\text{fail}<t>A=\text{wait}_t\mathbf{;}A\]
\item
  Failover to certain failure only cuts the latency curve:
  \[A<t>\text{fail}=\text{cut}_{t}A\]
\item
  Certain success ignores deadline: \[1_Q<t>A=1_Q \text{ when } t>0\]
\item
  Failover with no time left means discarding initial operation:
  \[A<0>B=B\]
\item
  When deadline is not shorter than maximum latency of left argument, it
  is similar to alternative, with extra delay before second argument:
  \[A<t>B=A ∨\text{wait}(t)B \text{ when } t>d(A)\]
\end{itemize}

\begin{enumerate}
\def\labelenumi{\arabic{enumi}.}
\setcounter{enumi}{4}
\item
  Retransmission without rejection of previous messages: \(A<t>>B\),
  when we have a little different algebraic properties with \emph{uncut}
  left argument: \[A<t>B=A ∨ \text{wait}_t\mathbf{;}B\] \[A<0>B=A ∨ B\]
  \[A<t>\text{fail}=A\] \[\text{fail}<t>A=A\]
\item
  Some approaches {[}3{]} propose using operator \(A⇆_{p} B\) for
  probabilistic choice between scenarios \(A\) with probability \(p\),
  and \(B\) with probability \(1-p\). In this work we assume the only
  way to get non-determinism is due to latency, for example if protocols
  use \textbf{unique agency property} like those used in Cardano network
  layer {[}{\textbf{???}}{]}.
\end{enumerate}

\hypertarget{abstracting-over-implementation}{%
\subsection{Abstracting over
implementation}\label{abstracting-over-implementation}}

We can encapsulate basic operations with
\passthrough{\lstinline!TimeToCompletion!} class, describing interface
that will be used both for latency distributions and their
approximations:

\begin{lstlisting}[language=Haskell]
class TimeToCompletion ttc where
  firstToFinish :: ttc -> ttc -> ttc
  lastToFinish  :: ttc -> ttc -> ttc
  after         :: ttc -> ttc -> ttc
  delay         :: Delay -> ttc
  allLost       :: ttc
  noDelay       :: ttc
  noDelay        = delay 0
\end{lstlisting}

\begin{lstlisting}[language=Haskell]
infixr 7 `after`
infixr 5 `firstToFinish`
infixr 5 `lastToFinish`

instance Probability            a
      => TimeToCompletion
           (LatencyDistribution a) where
  firstToFinish = firstToFinishLD
  lastToFinish  = lastToFinishLD
  after         = afterLD
  delay         = delayLD
  allLost       = allLostLD
  noDelay       = noDelayLD
\end{lstlisting}

\hypertarget{general-treatment-of-completion-distribution-over-time}{%
\subsubsection{General treatment of completion distribution over
time}\label{general-treatment-of-completion-distribution-over-time}}

Whether might aim for minimum delay distribution of message over a given
connection \(∆Q(t)\) , minimum time of propagation of the message over
entire network (\(∆R(t)\), reachability), we still have a distribution
of completion distribution over time with standard operations.

We will need a standard library for treating these to speed up our
computations.

We can also define a mathematical ring of (probability, delay) pairs.

Note that \passthrough{\lstinline!LatencyDistribution!} is a modulus
over ring R with \passthrough{\lstinline!after!} as multiplication, and
\passthrough{\lstinline!whicheverIsFaster!} as addition. Then
\passthrough{\lstinline!noDelay!} is neutral element of multiplication
(unit or one), and \passthrough{\lstinline!allLost!} is neutral element
of addition.\footnote{This field definition will be used for
  multiplication of connection matrices.} Note that both of these binary
operators give also rise to two almost-scalar multiplication operators:

\begin{lstlisting}[language=Haskell]
scaleProbability :: Probability         a
                 =>                     a
                 -> LatencyDistribution a
                 -> LatencyDistribution a
scaleProbability a = after $ preserved a

scaleDelay  :: Probability         a
            => Delay
            -> LatencyDistribution a
            -> LatencyDistribution a
scaleDelay t = after $ delayLD t

delayLD :: Probability a
        => Delay
        -> LatencyDistribution a
delayLD n = LatencyDistribution
        $ Series
        $ [0.0 | _ <- [(0::Delay)..n-1]] <> [1.0]
\end{lstlisting}

To compare distributions represented by series of approximate values we
need approximate equality:

\begin{lstlisting}[language=Haskell]
instance (Metric                a
         ,Num                   a
         ,Real                  a)
      => Metric
           (LatencyDistribution a) where
  LatencyDistribution l
    `distance`
      LatencyDistribution m =
        realToFrac $ sum $ fmap (^2) $ l-m
  similarityThreshold = 1/1000
\end{lstlisting}

Choosing \passthrough{\lstinline!0.001!} as similarity threshold (should
depend on number of samples)

\begin{lstlisting}[language=Haskell]
instance Unit a
      => Unit (LatencyDistribution a) where
  unitE = LatencyDistribution (Series [unitE])

instance Null a
      => Null (LatencyDistribution a) where
  nullE = LatencyDistribution (Series [nullE])
\end{lstlisting}

\hypertarget{bounds-on-distributions}{%
\subsection{Bounds on distributions}\label{bounds-on-distributions}}

Note that we can define bounds on
\passthrough{\lstinline!LatencyDistribution!} that behave like functors
over basic operations from \passthrough{\lstinline!TimeToCompletion!}
class.

\begin{itemize}
\tightlist
\item
  Upper bound on distribution is the \passthrough{\lstinline!Latest!}
  possible time\footnote{Here \passthrough{\lstinline!liftBinOp!} is for
    lifting an operator to a newtype.}:
\end{itemize}

\begin{lstlisting}[language=Haskell]
newtype Latest =
        Latest { unLatest :: SometimeOrNever }
  deriving (Eq, Ord, Show)

newtype SometimeOrNever =
        SometimeOrNever
          { unSometimeOrNever :: Maybe Delay }
  deriving (Eq)

instance Show SometimeOrNever where
  showsPrec _     Never       = ("Never"++)
  showsPrec prec (Sometime t) =
      showParen (prec>app_prec) $
        ("Sometime "++) . showsPrec (app_prec+1) t
    where
      app_prec = 10

pattern Never      = SometimeOrNever Nothing
pattern Sometime t = SometimeOrNever (Just t)

instance Ord SometimeOrNever where
  Never      `compare` Never      = EQ
  Never      `compare` Sometime _ = GT
  Sometime _ `compare` Never      = LT
  Sometime t `compare` Sometime u = t `compare` u

latest :: Probability         a
       => LatencyDistribution a
       -> Latest
latest ((<1.0) . sum . pdf -> True  ) = Latest Never
latest (last . unSeries . pdf -> 0.0) =
  error ("Canonical LatencyDistribution "
     <>  "should always end with non-zero value")
latest  x                             =
    Latest . Sometime . Delay . (-1+)
  . length . unSeries . pdf $ x

onLatest = liftBinOp unLatest Latest

instance TimeToCompletion Latest where
  firstToFinish = onLatest min
  lastToFinish  = onLatest max
  after         =
    liftBinOp (unSometimeOrNever . unLatest)
              (Latest . SometimeOrNever)
              (liftM2 (+))
  delay         = Latest . Sometime
  allLost       = Latest   Never
\end{lstlisting}

\begin{itemize}
\tightlist
\item
  Lower bound on distribution is the \passthrough{\lstinline!Earliest!}
  possible time:
\end{itemize}

\begin{lstlisting}[language=Haskell]
newtype Earliest =
        Earliest
          { unEarliest :: SometimeOrNever }
  deriving (Eq, Ord, Show)
earliest :: Probability         a
         => LatencyDistribution a
         -> Earliest
earliest [0.0]  = Earliest Never
earliest [_]    = Earliest $ Sometime 0
earliest (last . unSeries . pdf -> 0.0) =
  error ("Canonical LatencyDistribution "
     <>  "should always end with non-zero value")
earliest  other = Earliest . Sometime
                . Delay . max 0 . length
                . takeWhile (0==) . unSeries
                . pdf $ other

onEarliest     = liftBinOp unEarliest Earliest

instance TimeToCompletion Earliest where
  firstToFinish = onEarliest min
  lastToFinish  = onEarliest max
  after         =
    liftBinOp (unSometimeOrNever . unEarliest)
              (Earliest . SometimeOrNever)
              (liftM2 (+))
  delay         = Earliest . Sometime
  allLost       = Earliest Never
\end{lstlisting}

These estimates have the property that we can easily compute the same
operations on estimates, without really computing the full
\passthrough{\lstinline!LatencyDistribution!}.

\hypertarget{failure-models}{%
\section{Failure models}\label{failure-models}}

\hypertarget{transient-failures}{%
\subsection{Transient failures}\label{transient-failures}}

Up to now, we used probability distributions and their discretizations
in order to model \emph{transient failures}:

\begin{itemize}
\tightlist
\item
  congestion leading to dropped packages
\item
  transmission errors leading to dropped messages
\end{itemize}

\hypertarget{persistent-failures}{%
\subsection{Persistent failures}\label{persistent-failures}}

In order to deal with \emph{persistent failures}, we would need a notion
of failure that is not independentent: when we see persistent failure,
it is likely that we will see it again during retransmissions.

Since persistent failures are likely to continue, we can easily model
them as a change of network topology: some connection latencies will be
reset to zero, when persistent failure between two nodes occurs.

We can easily accomodate it as a new layer over our model of the
network: we will assign probabilities to each link about how often the
persistent failure occurs, and compute with these in mind.

\hypertarget{representing-networks}{%
\section{Representing networks}\label{representing-networks}}

Adjacency matrix is classic representation of network graph, where
\(i\)-th row corresponds to outbound edges of node \(i\), and \(j\)-th
column corresponds to inbound edges of node \(j\). So \(A_{i,j}\) is
\(1\) when edge is connected, and \(0\) if edge is not connected.

It is common to store only upper triangular part of the matrix, since:

\begin{itemize}
\tightlist
\item
  it is symmetric for undirected graphs
\item
  it should have \(1\) on the diagonal, since every node is connected to
  itself.
\end{itemize}

We use this trick to avoid double counting routes with different
directions.

So \emph{network connectivity matrix} is:

\begin{itemize}
\tightlist
\item
  having units on the diagonal
\item
  having connectivitity information between nodes \(i\), and \(j\), for
  \(j>i\) in element \(a_{i,j}\).
\end{itemize}

Generalizing this to ΔQ-matrices we might be interesting in:

\begin{itemize}
\tightlist
\item
  whether \(A^n\) correctly mimicks shortest path between nodes
  (\passthrough{\lstinline!Earliest!})
\item
  whether \(A^n\) correctly keeps paths shorter than \(n\)
\item
  for a strongly connected graph there should exist
  \(n≤\mathop{dim}(A)\), such that \(A^n\) is having non-null elements
  on an upper triangular section.
\end{itemize}

More rigorous formulation is: \[ \begin{array}{rcl}
R_0(A) & = & \mathit{1} \\
R_n(A) & = & R_{n-1}(A)*A \\
\end{array}
\] Where:

\begin{itemize}
\tightlist
\item
  \(\mathit{1}\) or \(\mathop{Id}\) denotes a unit adjacency matrix,
  that is matrix where every node is connected with itself but none
  else. And these connections have no delay at all.
\item
  \(A\) is connection matrix as defined above, and distribution for a
  transmission from a single packet from \(i\)-th to \(j\)-th node. For
  pre-established TCP this matrix should be symmetric.
\end{itemize}

Our key metric would be diffusion or reachability time of the network
\(∆R(t)\), which is conditioned by quality of connection curves
\(∆Q(t)\) and the structure network graph.

Later in this section, we discuss on how \(∆R(t)\) encompasses other
plausible performance conditions on the network.

This gives us interesting example of using matrix method over a modulus,
and not a group.

\hypertarget{reachability-of-network-broadcast-or-rt}{%
\subsubsection{Reachability of network broadcast or
∆R(t)}\label{reachability-of-network-broadcast-or-rt}}

Reachability curve \(∆R(t)\) or \emph{diffusion time} is plotted as
distribution of event where all nodes are reached by broadcast from
committee node, against time. We want to sum the curve for all possible
core nodes, by picking a random core node.

Area under the curve would tell us the overall quality of the network.
When curve reaches 100\% rate, then we have a strongly connected
network, which should be eventually always the case after necessary
reconfigurations.

\emph{Note that when running experiments on multiple networks, we will
need to indicate when we show average statistics for multiple networks,
and when we show a statistic for a single network.}

\hypertarget{description-of-network-connectivity-graph-in-terms-of-q}{%
\subsubsection{Description of network connectivity graph in terms of
∆Q}\label{description-of-network-connectivity-graph-in-terms-of-q}}

Traditional way of describing graphs is by adjacency matrix, where 0
means there is no edge, and 1 means that there is active edge.

We may generalize it to unreliable network connections described above,
by using \(ΔQ\) instead of binary.

So for each value diagonal, the network connection matrix \(A\) will be
\passthrough{\lstinline!noDelay!}, and \(A_{i j}\) will represent the
connection quality for messages sent from node \(i\) to node \(j\).

That allows us to generalize typical graph algorithms executed to
algorithms executed on network matrices:

\begin{enumerate}
\def\labelenumi{\arabic{enumi}.}
\tightlist
\item
  If series \(R_n(A)\) converges to matrix of non-zero
  (non-\passthrough{\lstinline!allLost!}) values \emph{in all cells} in
  a finite number of steps, we consider graph to be \emph{strongly
  connected} {[}9{]}. Matrix multiplication follows uses
  \((\mathbf{;},∨)\) instead of \((*,+)\). So sequential composition
  \passthrough{\lstinline!afterLD!} in place of multiplication, and
  alternative selection \passthrough{\lstinline!firstToFinish!} in place
  of addition.
\end{enumerate}

When it exists, limit of the series \(R_n(A)=A^n\) is called \(A^{*}\).

In case of non-zero delays, outside diagonal, we may also consider
convergence for delays up to \(t\). NOTE: \emph{We need to add estimate
of convergence for cutoff time \(t\) and number of iterations \(n\),
provided that least delay is in some relation to \(n*t\).}

Also note that this series converges to \(ΔQ\) on a single shortest path
between each two nodes. That means that we may call this matrix
\(R_{min}(t)\), or optimal diffusion matrix.

\begin{lstlisting}[language=Haskell]
optimalConnections  ::
    (Probability                     a
    ,KnownNat n
    ,Real                            a
    ,Metric                          a)
  => SMatrix  n (LatencyDistribution a)
  -> SMatrix  n (LatencyDistribution a)
optimalConnections a =
  converges (fromIntegral $ size a)
                          (|*|a) a
\end{lstlisting}

Of course, this requires a reasonable approximate metric
\passthrough{\lstinline!\~\~!} and definition of convergence:

\begin{lstlisting}[language=Haskell, label=converges]
converges ::  Metric r
          =>  Int -- ^ max number of steps
          -> (r -> r)
          ->  r
          ->  r
converges 0      step r =
  error "Solution did not converge"
converges aLimit step r | r ~~ r' =
    then                           r
    else converges (pred aLimit) step r'
  where
    r' = step r
\end{lstlisting}

We will use this to define \emph{path with shortest ΔQ}. It corresponds
to the situation where all nodes broadcast value from any starting point
\(i\) for the duration of \(n\) retransmissions.\footnote{That we do not
  reduce loss over remainder yet?}

\begin{enumerate}
\def\labelenumi{\arabic{enumi}.}
\setcounter{enumi}{1}
\item
  Considering two nodes we may consider delay introduced by
  retransmissions in naive miniprotocol:

  \begin{itemize}
  \tightlist
  \item
    we have two nodes \emph{sender} and \emph{receiver}
  \item
    \emph{sender} sends message once per period equal maximum network
    latency \passthrough{\lstinline!deadline!}
  \item
    the message is \emph{resent} if \emph{receiver} fails to send back
    confirmation of receipt \ldots{}
  \end{itemize}

  Assuming latency of the connection \(l\), and timeout \(t>d(l)\), we
  get simple solution: \[ \mu{}X.l\mathbf{;}l\mathbf{;}X \]
\item
  We need to consider further examples of how our metrics react to
  issues detected by typical graph algorithms.
\end{enumerate}

We define a matrix multiplication that uses
\passthrough{\lstinline!firstToFinish!} in place of addition and
\passthrough{\lstinline!after!} in place of multiplication.

\begin{lstlisting}[language=Haskell]
(|*|) :: (Probability                     a
         ,KnownNat n                       )
      =>  SMatrix  n (LatencyDistribution a)
      ->  SMatrix  n (LatencyDistribution a)
      ->  SMatrix  n (LatencyDistribution a)
(|*|)  = sMatMult firstToFinish after
\end{lstlisting}

Note that to measure convergence of the process, we need a notion of
distance between two matrices.

Here, we use Frobenius distance between matrices, parametrized by the
notion of distance between any two matrix elements.

\begin{lstlisting}[language=Haskell]
instance (Metric            a
         ,KnownNat        n  )
      =>  Metric (SMatrix n a) where
  a `distance` b =
      sqrt $
        sum [square ((a !(i,k)) `distance` (b ! (i,k)))
               | i <- allUpTo, k<-allUpTo]
    where
      square x = x*x
  similarityThreshold = similarityThreshold @a
\end{lstlisting}

\hypertarget{histogramming}{%
\section{Histogramming}\label{histogramming}}

To provide histograms of average number of nodes reached by the
broadcast, we need to define additional operations:

\begin{itemize}
\tightlist
\item
  sum of mutually exclusive events
\item
  K-out-of-N synchronization
\end{itemize}

\hypertarget{sum-of-mutually-exclusive-events}{%
\subsection{Sum of mutually exclusive
events}\label{sum-of-mutually-exclusive-events}}

First we need to define a precise sum of two events that are mutually
exclusive \(⊕\). That is different from
\passthrough{\lstinline!firstToFinish!} which assumes that they are
mutually independent.

\begin{lstlisting}[language=Haskell]
infixl 5 `exSum` -- like addition
class ExclusiveSum a where
  exAdd :: a -> a -> a

exSum :: ExclusiveSum a => [a] -> a
exSum  = foldr1 exAdd

instance ExclusiveSum ApproximateProbability where
  exAdd = (+)

instance ExclusiveSum IdealizedProbability where
  exAdd = (+)
\end{lstlisting}

Given a definition of exclusive sum for single events, and existence
neutral element of addition, we can easily expand the definition to the
latency distributions:

\begin{lstlisting}
instance ExclusiveSum a
      => ExclusiveSum (Series a) where
  Series a `exAdd` Series b = Series
    $ zipWithExpanding exAdd a b

instance ExclusiveSum                      a
      => ExclusiveSum (LatencyDistribution a) where
  LatencyDistribution     a `exAdd`
    LatencyDistribution             b =
    LatencyDistribution $ a `exAdd` b
\end{lstlisting}

\hypertarget{k-out-of-n-synchronization-of-series-of-events-a_k.}{%
\subsection{\texorpdfstring{\emph{K-out-of-N synchronization} of series
of events
\(a_k\).}{K-out-of-N synchronization of series of events a\_k.}}\label{k-out-of-n-synchronization-of-series-of-events-a_k.}}

For histogramming a fraction of events that have been delivered within
given time, we use generalization of recursive formula \(\binom{n}{k}\).

\[ F_{\binom{a_n}{k}}(t) = a_n ∧ F_{\binom{a_{n-1}}{k-1}}(t)
                         ⊕ (\overline{a_n} ∧ F_{\binom{a_{n-1}}{k}}(t)) \]

Here \(a_{n-1}\) is a finite series without its last term (ending at
index \(n-1\)).

It is more convenient to treat \(F_{\binom{a_n}{k}}(t)\) as a
\passthrough{\lstinline!Series!} with indices ranging over \(k\):
\(F_{\binom{a_n}{...}}(t)\). Then we see the following equation:
\[F_{\binom{a_{n}}{...}} = [\overline{a_n}, a_n] ⊛ F_{\binom{a_{n-1}}{...}} \]
Where:

\begin{itemize}
\tightlist
\item
  \(⊛\) is convolution
\item
  \([\overline{a_n}, a_n]\) is two element series having complement of
  \(a_n\) as a first term, and \(a_n\) as a second term.
\end{itemize}

We implement it as a series \passthrough{\lstinline!k\_of\_n!} with
parameter given as series \(a_n\), and indices ranging over \(k\):

\begin{lstlisting}[language=Haskell]
kOutOfN :: (TimeToCompletion a
           ,ExclusiveSum     a
           ,Complement       a
           ) => Series       a
             -> Series       a
kOutOfN (Series []    ) =
  error "kOutOfN of empty series"
kOutOfN (Series [x]   ) = [complement x, x]
kOutOfN (Series (x:xs)) = [x] `convolution`
                          Series xs
  where
    convolution = convolve_ exAdd lastToFinish
             `on` kOutOfN
\end{lstlisting}

\hypertarget{fraction-of-reached-nodes}{%
\subsubsection{Fraction of reached
nodes}\label{fraction-of-reached-nodes}}

Now, for a connection matrix \(A\), each row corresponds to a vector of
latency distribution for individual nodes. Naturally source node is
indicated as a \emph{unit} on a diagonal. Now we can use
\passthrough{\lstinline!kOutOfN!} to transform the series corresponding
to a single row vector into a distribution of latencies for reaching
\emph{k-out-of-n} nodes. Note that this new series will have
\emph{indices} corresponding to \emph{number of nodes reached} instead
of node indices:

\begin{lstlisting}[language=Haskell]
nodesReached :: (Probability  a
                ,ExclusiveSum a)
             => Series (LatencyDistribution a)
             -> Series (LatencyDistribution a)
nodesReached  = kOutOfN
\end{lstlisting}

For this we need to define \passthrough{\lstinline!complement!} for
\passthrough{\lstinline!LatencyDistribution!}:

\begin{lstlisting}[language=Haskell]
instance Complement                      a
      => Complement (LatencyDistribution a) where
  complement (LatencyDistribution s) =
    LatencyDistribution
      ( complement <$> s )
\end{lstlisting}

\hypertarget{averaging-broadcast-from-different-nodes}{%
\subsection{Averaging broadcast from different
nodes}\label{averaging-broadcast-from-different-nodes}}

Given that we have a connection matrix \(A\) of broadcast iterated \(n\)
times, we might want histogram of distribution of a fraction of nodes
reached for a random selection of source node.

We can perform this averaging with exclusive sum operator, pointwise
division of elements by the number of distributions summed:

\begin{lstlisting}[language=Haskell]
averageKOutOfN  :: (KnownNat     n
                   ,ExclusiveSum                  a
                   ,Probability                   a
                   ,Show                          a)
                => SMatrix n (LatencyDistribution a)
                -> Series    (LatencyDistribution a)
averageKOutOfN m = average
  (nodesReached . Series <$> rows m)

average :: (ExclusiveSum                a
           ,Probability                 a
           ,Show                        a)
        => [Series (LatencyDistribution a)]
        ->  Series (LatencyDistribution a)
average aList =  scaleLD
             <$> exSum aList
  where
    scaleLD :: Probability         a
            => LatencyDistribution a
            -> LatencyDistribution a
    scaleLD = scaleProbability
             (1/fromIntegral (length aList))
\end{lstlisting}

\hypertarget{summary}{%
\section{Summary}\label{summary}}

We have shown how few lines of Haskell code can be used to accurately
model network latency and get n-hop approximations of packet
propagation.

It turns out they can also be used to model task completion
distributions for a all-around estimation of software completion time.

\hypertarget{what-for-capacity-limited-networks}{%
\subsection{What for capacity-limited
networks?}\label{what-for-capacity-limited-networks}}

It turns out that most of the real network traffic is latency limited
and focused on \emph{mice} connections: that is connections that never
have bandwidth-latency product that would be greater than 12kB*s.

That means that our approximation is actually useful for most of the
flows in real networks, even though the real connections have limited
capacity!

\hypertarget{relation-of-latency-to-other-plausible-metrics-of-network-performance}{%
\subsection{Relation of latency to other plausible metrics of network
performance}\label{relation-of-latency-to-other-plausible-metrics-of-network-performance}}

One can imagine other key properties that network must satisfy:

\begin{itemize}
\item
  That absent permanent failures, network will reach full connectivity.
  That corresponds to the situation where given ∆Q(t) iCDF ultimately
  reaches 100\% delivery probability for some delay, ∆R(t) will also
  always reach 100\%. \emph{Moreover ∆R(t) metric allows to put deadline
  for reaching full connectivity in a convenient way.}
\item
  That given a fixed limit on rate of nodes joining and leaving the
  network, we also will have deadline on when ∆R(t) reaches a fixed
  delivery rate \(R_{SLA}\) within deadline \(t_{SLA}\).
\item
  That given conditions on minimum average quality of connections
  \(∆Q(t)\), and fixed rate of adversary nodes \(r_{adv}\) we can still
  guarantee networks reaches reachability \(R_{SLA}\).
\item
  That there are conditions for which \(∆R(t)\) always reaches almost
  optimal reachability defined by given ratio \(o\in{}(0.9,1.0)\), such
  that \(∆R(o*t) \ge max(∆R_{optimal}(t)/o)\). In other words: there is
  a deadline \(o\)-times longer than time to reach optimal reachability
  in an optimal network, we reach connectivity of no less than
  \(o\)-times connectivity of the optimal network.
\end{itemize}

\hypertarget{interesting-properties}{%
\subsection{Interesting properties}\label{interesting-properties}}

We note that moduli representing latency distributions have properties
that allow for efficient estimation by bounds that conform to the same
laws.

Square matrices of these distributions or their estimations can be used
to estimate network propagation and reachability properties.

That makes for an interesting class of algebras that can be used as a
demonstration of moduli to undergraduates, and also allows to introduce
them to latency-limited performance which is characteristic to most of
modern internetworking.

\hypertarget{future-work}{%
\subsection{Future work}\label{future-work}}

We would like to apply these methods of latency estimation to modelling
a most adverse scenarios: when hostile adversary aims to issue
denial-of-service attack by delaying network packets{[}1{]}.

\hypertarget{bibliography}{%
\section*{Bibliography}\label{bibliography}}
\addcontentsline{toc}{section}{Bibliography}

\hypertarget{refs}{}
\leavevmode\hypertarget{ref-ProgrammingSatan}{}%
{[}1{]} Anderson, C. et al. 2005. Towards type inference for javascript.
\emph{ECOOP 2005 - object-oriented programming} (Berlin, Heidelberg,
2005), 428--452.

\leavevmode\hypertarget{ref-mice}{}%
{[}2{]} Azzana, Y. et al. 2009. Adaptive algorithms for identifying
large flows in ip traffic.

\leavevmode\hypertarget{ref-bradley}{}%
{[}3{]} Bradley, J.T. 1999. \emph{Towards reliable modelling with
stochastic process algebras}. Department of Computer Science, University
of Bristol.

\leavevmode\hypertarget{ref-NetworkRobustness}{}%
{[}4{]} Ellens, W. and Kooij, R.E. 2013. Graph measures and network
robustness. \emph{CoRR}. abs/1311.5064, (2013).

\leavevmode\hypertarget{ref-EffectiveGraphResistance}{}%
{[}5{]} Ellens, W. et al. 2011. Effective graph resistance. \emph{Linear
Algebra and its Applications}. 435, 10 (2011), 2491--2506.
DOI:\url{https://doi.org/https://doi.org/10.1016/j.laa.2011.02.024}.

\leavevmode\hypertarget{ref-network-calculus}{}%
{[}6{]} Jean-Yves Le Boudec; Thiran, Patrick 2001. Network Calculus: A
Theory of Deterministic Queuing Systems for the Internet. Lecture Notes
in Computer Science. 2050.

\leavevmode\hypertarget{ref-PowerSeries}{}%
{[}7{]} McIlroy, M.D. 1999. Power series, power serious. \emph{Journal
of Functional Programming}. (1999), 323--335.

\leavevmode\hypertarget{ref-ProgramAnalysis}{}%
{[}8{]} Nielson, F. et al. 1999. \emph{Principles of program analysis}.

\leavevmode\hypertarget{ref-GeneralMethodOfShortestPaths}{}%
{[}9{]} O'Connor, R. 2011. A very general method of computing shortest
paths. \emph{Russell O'Connor's Blog}. http://r6.ca/.

\leavevmode\hypertarget{ref-wiki:backwardFiniteDifference}{}%
{[}10{]} Wikipedia contributors 2019. Finite difference --- Wikipedia,
the free encyclopedia.
\url{https://en.wikipedia.org/w/index.php?title=Finite_difference\&oldid=903566420}.

\leavevmode\hypertarget{ref-wiki:antidifference}{}%
{[}11{]} Wikipedia contributors 2019. Indefinite sum --- Wikipedia, the
free encyclopedia.
\url{https://en.wikipedia.org/w/index.php?title=Indefinite_sum\&oldid=900662530}.

\hypertarget{glossary}{%
\section*{Glossary}\label{glossary}}
\addcontentsline{toc}{section}{Glossary}

\begin{itemize}
\tightlist
\item
  \(t∈\mathcal{T}\) - time since sending the message, or initiating a
  process
\item
  \(∆Q(t)\) - response rate of a single connection after time \(t\)
  (chance that message was received until time \(t\))
\item
  \(∆R(t)\) - completion rate of broadcast to entire network (rate of
  nodes expected to have seen the result until time \(t\))
\item
  \(\epsilon{}\) - rate of packets that are either dropped or arrive
  after latest reasonable deadline we chose
\end{itemize}

\bibliography{Latency.bib}

\end{document}